**Tailoring Vanadium Dioxide Film Orientation using Nanosheets: A Combined Microscopy, Diffraction, Transport and Soft X-ray in Transmission Study**

*Phu Tran Phong Le[1], Kevin Hofhuis [1], Dr. Abhi Rana [1], Prof. Mark Huijben [1], Prof. Hans Hilgenkamp [1], Prof. G. Rijnders [1], Prof. A. ten Elshof [1], Prof. Gertjan Koster[\*][1], Dr. Nicolas Gauquelin[\*][1,2], Gunnar Lumbeeck [2], Dr. Christian Schlüßler-Langeheine [3], Dr. Horia Popescu [4], F. Fortuna [5], Steef Smit [6], Xanthe H. Verbeek [6], Georgios Araizi-Kanoutas [6], Dr. Shrawan Mishra [6,7], Dr. Igor Vakivskyi[8], Prof. Hermann A. Dürr [6,8], Prof. Mark S. Golden[\*][6]*

1
MESA+ institute for nanotechnology, University ofTwente, PObox 217, Enschede, The Netherlands.
2 Electron Microscopy for Materials Science (EMAT), University of Antwerp, 2020 Antwerp, Belgium
3 Helmholtz-Zentrum Berlin für Materialien und Energie, BESSY II, Albert-Einstein-Str. 15 12489 Berlin, Germany
4 Synchrotron SOLEIL, L'Orme des Merisiers Saint-Aubin, BP 48 91192 Gif-sur-Yvette Cedex, France
5 CSNSM, Université Paris-Sud and CNRS/IN2P3, Bâtiments 104 et 108, 91405 Orsay cedex, France
6 Van der Waals-Zeeman Institute, Institute of Physics, Science Park 904, 1098 XH, Amsterdam, The Netherlands
7 School of Material Science and Technology, Indian Institute of Technology (BHU), Varanasi 221005, India
8 Department of Physics and Astronomy, Uppsala University, Box 516 751 20 Uppsala, Sweden

Corresponding authors:	G.Koster@utwente.nl

Nicolas.Gauquelin@uantwerpen.be

M.S.Golden@uva.nl

Abstract:

VO$_2$ is a much-discussed material for oxide electronics and neuromorphic computing applications. Here, heteroepitaxy of vanadium dioxide (VO$_2$) was realized on top of oxide nanosheets that cover either the amorphous silicon dioxide surfaces of Si substrates or X-ray transparent silicon nitride membranes. The out-of-plane orientation of the VO$_2$ thin films was controlled at will between (011)$_{M1}$/(110)$_R$ and (-402)$_{M1}$/(002)$_R$ by coating the bulk substrates with $Ti_{0.87}O_2$ and $NbWO_6$ nanosheets, respectively, prior to $VO_2$ growth. Temperature-dependent X-ray diffraction and automated crystal orientation mapping in microprobe TEM mode (ACOM-TEM) characterized the high phase purity, the crystallographic and orientational properties of the $VO_2$ films. Transport measurements and soft X-ray absorption in transmission are used to probe the $VO_2$ metal-insulator transition, showing results of a quality equal to those from epitaxial films on bulk single-crystal substrates. Successful local manipulation of two different $VO_2$ orientations on a single substrate is demonstrated using $VO_2$ grown on lithographically-patterned lines of $Ti_{0.87}O_2$ and $NbWO_6$ nanosheets investigated by electron backscatter diffraction. Finally, the excellent suitability of these nanosheet-templated $VO_2$ films for advanced lens-less imaging of the metal-insulator transition using coherent soft X-rays is discussed.

**1. Introduction**

Vanadium dioxide ($VO_2$) has been drawing attention since the discovery of its metal-insulator transition (MIT), signaled by a several orders of magnitude resistivity change close to 340 K[1,2]. Given these remarkable properties, it may not be a surprise that $VO_2$ is a leading candidate material for the development of oxide devices for both low power electronics (high off-resistance), either in a more conventional field effect type of device or alternatively, neuromorphic electronic architectures[3,4] as a memristive material. It has been known that the MIT occurs alongside an abrupt, first-order structural phase transformation from a metallic, tetragonal rutile (R) phase ($P4_2/mnm$), to an insulating, monoclinic (M1) phase ($P2_1/C$). Recent work[5] points out that this transition is preceded by a purely electronic softening of Coulomb correlations within the V-V singlet dimers that characterize the insulating state, setting the energy scale for driving the near-room-temperature insulator-metal transition in this paradigm complex, correlated oxide.

Up to now, most studies of epitaxial $VO_2$ thin films have used $Al_2O_3$ and $TiO_2$ single crystal substrates to control film orientation[6,7], bringing along challenges of cost, limited size and incompatibility with the current Si-based technology for future $VO_2$-based devices. Direct deposition of $VO_2$ on glass or Si substrates with a native amorphous silicon dioxide layer leads to a polycrystalline film with predominant $(011)_{M1}$ orientation[8], whereas $VO_2$ is favorably grown $(010)_{M1}$-oriented when a buffer layer of Pt(111) is used on a Si substrate[9]. Epitaxial growth of $VO_2$ with $(010)_{M1}$ orientation is possible on epitaxial layers of yttria-stabilized zirconia - YSZ(001) - on Si(001)[10,11].

Ideally, one would wish for direct control over $VO_2$ film orientation on Si or even on arbitrary substrates at will, without any concessions being made on the $VO_2$ film quality. Oriented growth is not only an important enabler for the fundamental study of the mechanism of the $VO_2$ MIT[12], but also for potential applications for next-generation transistors[13], memory metamaterials[14], sensors[15], and novel hydrogen storage technology[16]. Recently, various metal oxide films have been successfully grown on glass and Si substrates using epitaxy on so-called *oxide nanosheets*[17–20]. Oxide nanosheets are essentially two dimensional (2D) single crystals with a thickness of a few nanometers or less, and lateral size in the micrometer range. They can be made spanning a wide range of crystal lattices and 2D structural symmetries [21], allowing for new possibilities to tailor the important structural parameters and properties of thin films on arbitrary – and thus also technologically-relevant - bulk substrates. Successful implementation of nanosheets in fact means that the choice of the bulk substrate becomes a free parameter that can enter the engineering cycle of each individual application.

In the research presented here, $Ti_{0.87}O_2$ (TO) and $NbWO_6$ (NWO) nanosheets have been identified as being ideal templates for the orientation of thin films of the important complex oxide $VO_2$ on varying substrates. Monolayers of nanosheets were deposited on Si substrates and alternatively on 20nm thick $Si_3N_4$ TEM grids using the Langmuir-Blodgett (LB) method. Then, utilizing pulsed laser deposition (PLD), single-phase $VO_2$ thin films were grown epitaxially on both TO and NWO nanosheets with $(011)_{M1}$ [$(110)_R$] and $(-402)_{M1}$ [$(002)_R$] out-of-plane orientation of the low temperature monoclinic M1 phase [high-T rutile phase], respectively. The high structural and orientational quality of the VO$_2$ made possible by the nanosheet epitaxy was proven using TEM and X-ray diffraction across the Mott MIT, as well as by electron backscatter diffraction (EBSD) studies. In addition, both transport and soft X-ray spectroscopic probes of the MIT showed data of excellent quality, matching those for VO$_2$ grown on bulk single-crystalline substrates. Importantly, the use of a nanosheet-coated Si$_3$N$_4$ membrane as a PLD substrate allowed soft X-ray absorption experiments to be carried out in the fully bulk-sensitive and highly direct transmission mode. Finally, the nanosheet-approach is shown to provide a high degree of control of the crystallographic orientation of the VO$_2$ film. This is illustrated by the use of a single film-growth run to generate two different orientations on a single substrate deterministically, by arranging both TO and NWO nanosheets in an alternating, stripe-like pattern using lithography.

## 2. Results and discussion

**Figure 1** shows a plan view of the TO and NWO nanosheet planes in panels (a) and (d), respectively. For the TO[NWO] nanosheets, the relevant 2D unit cell is highlighted in green[pink]. In panels (b) and (c), the relevant VO$_2$ planes are shown for the M1 and R structures with the 2D unit cell of the TO nanosheets superimposed. Likewise, in panels (e) and (f) the 2D unit cell of the NWO nanosheets is superimposed on the relevant planes in the M1 and R VO$_2$ phases. What these figures show, backed up by the data of Table 1, is that the (011)$_{M1}$ [(110)$_R$] and (-402)$_{M1}$ [(002)$_R$] out-of-plane orientations of M1[R] VO$_2$ are compatible with the TO and NWO nanosheet symmetry and lattice constants, respectively.

In the following, the reasoning and data that led to this conclusion are gone through in a step-by-step manner. A closer examination of Figure 1 shows that - considering the oxygen sub-lattices as the dominating structural entities driving potential epitaxy between the VO$_2$ and the nanosheets - the two distinct O-O distances of 2.90 Å and 2.84 Å for the VO$_2$ (011)$_{M1}$ plane are close to the O-O distance of 2.97 Å for the TO nanosheet in the *b* direction. In the *a* direction, there are also two distinct O-O distances for the VO$_2$ (011)$_{M1}$ plane, i.e. 3.38 Å and 4.51 Å, respectively, which are very different from the O-O distance of 3.82 Å for TO nanosheets. Highly relevant in this regard is the concept of domain matching epitaxy[22], which can be seen as a generalization of the more common lattice match epitaxy. In domain matching epitaxy, integral multiples of lattice planes match across the film–substrate interface, with the size of the domain (equaling integral multiples of planar spacing), determined by the degree of mismatch. This can yield epitaxy when mismatches exceed the usual 7-8% limit for regular lattice epitaxy[22].

If we consider the domain epitaxy approach, it can be seen that the slightly distorted (only 0.35° away from a right-angle) rectangular oxygen domain of 7.56 x 5.74 Å$^2$ for the VO$_2$ (011)$_{M1}$ plane is close to that of 7.64 x 5.94 Å$^2$ for the TO nanosheets. This results in a small domain mismatch of only 1% and 3.4% in the *a*- and *b*-directions at room temperature, respectively, suggesting that the M1 monoclinic phase of VO$_2$ (011)$_{M1}$ film can be stabilized by TO nanosheets at room temperature. At the growth temperature of 520°C, it is the rutile phase of VO$_2$ that is being formed[2], and analogous considerations show that as the (110)$_R$ plane is very closely related to the room temperature (011)$_{M1}$ plane of VO$_2$ a similar domain epitaxial relationship will be at work.

In addition to the domain epitaxy ideas, we also point out that as the nanosheets are exfoliated layered materials without chemically active dangling bonds, there are only isotropic Coulomb and/or Van der Waals interactions between the growing film and the nanosheet surface. Consequently, lateral adatom-adatom interactions are significant drivers of the energetics of the early stages of (epitaxial) growth, thus helping to favor epitaxial growth also in the presence of relatively large lattice mismatch. In this manner, retention of epitaxy with lattice mismatch as high as 13% have been reported[19,23], supporting the ability of TO nanosheets to successfully template the (110)$_R$ growth of the rutile phase of VO$_2$.

We now turn to the NWO nanosheets. In this case, epitaxial growth of (-402)$_{M1}$ or (002)$_R$ VO$_2$ films are expected via straightforward lattice matching considerations. The 2D atomic structure of the (-402)$_{M1}$ plane is distorted away from a square planar symmetry (by 3° off the right-angle). The lattice mismatch between this plane and the nanosheets is 3.6% and 7.9% in the *a*- and *b*-directions, respectively, which – especially given the arguments above as regards the non-directional nature of the adatom-nanosheet interactions - we would expect to support epitaxial growth with the VO$_2$ film oriented in the [-402]$_{M1}$ direction at room temperature. For the NWO case, the high-T situation is simpler still, as the lattice mismatch between the high temperature (002)$_R$ plane and the NWO nanosheet is only 3.2%, with the same 2D square atomic structure present in both cases.

To summarize this section: detailed consideration of the oxygen sub-lattice-driven epitaxial relationships shows that for TO nanosheets, a combination of domain matching epitaxy and the weak adatom-nanosheet interactions should enable epitaxial growth of VO$_2$, with the out-of-plane orientation being (011)$_{M1}$ and (110)$_R$ at room temperature and elevated temperature, respectively. For the NWO nanosheets, the VO$_2$ orientation is expected to be (-402)$_{M1}$ and (002)$_R$.

Turning to the first step of the nanosheet-templated film growth process in practice, **Figure 2** shows AFM images of the morphology of the bare-nanosheets of TO and NWO in panels (a) and (b). It is clear that monolayers of TO and NWO nanosheets can be fabricated on Si substrates successfully with a surface coverage exceeding 95%. The lateral size of the individual nanosheets is 3-5 μm for both Ti$_{0.87}$O$_2$ and NWO$_6$, which is partly governed by the grain size of the parent layered crystals obtained by solid state reaction[24]. The exfoliated nanosheets were deposited using a LB-method to form a monolayer film on Si substrates and on Si$_3$N$_4$ membranes.

Using these nanosheet layers on Si$_3$N$_4$ TEM grids as substrates, thin films of VO$_2$ were grown using pulsed laser deposition. In order to determine the homogeneity and epitaxial quality of the VO$_2$ films with nm lateral spatial precision, orientation maps were recorded using TEM, the results of which are shown in Figure 2(c)-(e). The data were measured in the rutile VO$_2$ phase at a temperature of 423K. For the TO-templated system, a remarkably homogeneous $(110)_R$ orientation of the VO₂ film results (Figure 2(c)), and apart from a very few rogue patches, Figure 2(d) shows that the NWO nanosheets perform just as well, generating $(002)_R$ VO₂. These results are confirmed in the pole figures for these two rutile orientations shown in panel (e). With this nanoscopic confirmation of the excellent epitaxial growth on each individual nanosheet, the next step is a more global measure of the VO₂ structure, in both the rutile and monoclinic phases using X-ray diffraction.

**Figure 3** presents the XRD data measured below and above $T_{MIT}$ of VO₂ films grown using PLD at 520°C on monolayers of TO (panel (a)) and NWO nanosheets (panel(b)) on Si substrates. At 303 K, the peaks seen at 2θ values of 27.90° and 57.72° in Figure 3(a) are from the $(011)_{M1}$ and $(022)_{M1}$ VO₂ reflections, respectively. In Figure 3(b), the peak at 65.07° corresponds to the $(-402)_{M1}$ VO₂ reflection. At 403K - above the MIT temperature – peaks measured at 27.70°, 57.27° and 65.28° now correspond to the $(110)_R$, $(220)_R$ and $(002)_R$ Bragg peaks of rutile VO₂, respectively. These XRD data confirm the excellent orientational integrity of the VO₂ films at the macroscopic scale, confirming what was seen in the TEM data. This points towards the major role of the oxygen framework in the determination of the epitaxial relationships between both nanosheet systems and the VO₂ overlayer, as discussed earlier.

One of the motivations for choosing VO₂ for this study was its dual role as a model system for both understanding strongly correlated insulator-metal transitions, as well as its tunable/switchable large resistance change near room temperature[3]. Consequently, the transport behavior of VO₂ grown using PLD on nanosheet templates is of great interest, and these data are shown in **Figure 4**. Defining the midpoint of the transition in the resistance curve measured upon heating as the phase transition temperature, $T_{MIT}$, our data show transition temperatures close to the canonical value of 341 K for bulk VO₂ single crystals[1,2]. The $T_{MIT}$ of the films with out-of-plane $(110)_R$ texture (rutile c-axis in-plane) grown on TO nanosheets was 347 K, whereas for the out-of-plane $(001)_R$ textured film on NWO nanosheets $T_{MIT}$ was 332 K.

These values are in agreement with what has been found in the literature on the orientation dependence of $T_{MIT}$ on different bulk $TiO_2$ substrates[25].

The $T_{MIT}$ in $VO_2$ is related to the V-V distance along the rutile *c*-axis, which affects the orbital overlap and the metallicity in the rutile phase[26]. In the past, this had been studied as a function of film thickness and strain $(001)_R$ for films grown on single crystal substrates[25,27]. For example, under compressive strain, the $T_{MIT}$ of a 24 nm thick $(001)_R$ $VO_2$ film grown on $TiO_2$ decreased to 330 K, while it was the same as the bulk $VO_2$ value of 341 K for a fully-relaxed 74 nm thick film[27]. When the rutile c-axis is under tensile strain, similar to using $TiO_2$ (110) single crystal substrates, $T_{MIT}$ is seen to increase to 369 K[26]. Therefore, as $T_{MIT}$ for $VO_2$ grown on TO nanosheets is higher than that of strain-free, bulk $VO_2$, this indicates that the R-$VO_2$ *c*-axis (2.86 Å when unstrained) is under a degree of tensile strain in the [010] direction of the TO nanosheet which has $b$ = 2.97 Å. The situation is the other way around for the $VO_2$ grown on NWO nanosheets: here $T_{MIT}$ is lower, suggesting compressive strain along the rutile *c*-axis, as could be expected from the Poisson effect given the tensile strain along the [100] and [010] directions of $VO_2$ ($a$ = 4.53 Å) due to coupling to the NWO nanosheet ($a$ = 4.68 Å).

The MIT for the TO templated $VO_2$ shows a steep-sided hysteresis curve. In comparison, the data of Figure 4 shows that the $VO_2$ grown on NWO nanosheets displays a relatively broad MIT. A possible explanation for this observation could be related to the fact that the NWO-templated $VO_2$ possesses a surface roughness of order 10 nm, compared to 2 nm for the TO-templated $VO_2$ films (see Figure S1 for AFM data from the $VO_2$ films). As the MIT in $VO_2$ possesses a strongly percolative character[28], the increased roughness, on top of an increased density of grain boundaries due to inter-nanosheet boundaries[17] can be at least a partial explanation of the broader transition for the case of $VO_2$ grown on NWO nanosheet templates. In addition, the larger deviation from 90° bond angles of the 2D atomic structure of the $(-402)_{M1}$ plane in the NWO case compared to the $(011)_{M1}$ plane for TO-templated growth (see Figure 1) is also compatible with a larger domain/grain boundary contribution to the transport for the NWO-templated VO$_2$.

From the experimental data thus far, it is clear that the nanosheets provide an elegant and effective method to control the orientation of VO$_2$ films on two widely different bulk substrates. In order to both emphasize the added possibilities that freedom from 'hard' substrate epitaxy enables, as well as to illustrate how this can also be controlled on the micron scale, the next section reports a new strategy to manipulate different orientations of VO$_2$ on a single, arbitrary substrate. To do this, NWO nanosheets were lithographically line-patterned on top of a monolayer of TO nanosheets, with subsequent VO$_2$ growth on this structured template layer to demonstrate the local manipulation of the orientation of the resulting high-quality VO$_2$ film.

**Figure 5** presents the HR-SEM cross-sectional view of the line-patterned VO$_2$, showing in panel (a) that the film thickness was roughly 50 nm on both types of nanosheets. In addition, the HR-SEM plan-view (Figure 5(b)) clearly reveals the different surface morphologies alluded to earlier on either side of the nanosheet boundary, consistent with the AFM data of VO$_2$ films grown on 'single species' TO and NWO nanosheets (see Figure S1). In both situations the resulting VO$_2$ film has a smoother surface on TO than on NWO nanosheets.

In addition to SEM imaging, electron backscatter diffraction (EBSD) maps were recorded, providing crystallographic information on the VO$_2$ film on the lithographically patterned nanosheet layers, shown in Figure 5(c) and (d). As these data were recorded at room temperature, we use in the following the M1 film orientations (the corresponding R-phase orientation was shown in Figure 2). The inverse pole figure maps reveal the out-of-plane orientations (-402)$_{M1}$ (shown in gold color, becoming (002)$_R$ at elevated temperature) and (011)$_{M1}$ (shown in purple, becoming (110)$_R$ at elevated temperature). Taking a closer look at a domain on single-typed nanosheet regions, the four inverse pole figures of Figure 5(e) correspond to the four chosen regions indicated in Figure 5(c). These show the presence of a single out-of-plane orientation for each type of nanosheet. It is evident that the VO$_2$ film out-of-plane orientations follow the underlying nanosheets with high fidelity, down to the micron level as we designed.

The in-plane EBSD orientation map displayed in Figure 5(d) shows a random orientational distribution, resulting from the randomness of in-plane orientation of nanosheets during LB deposition. On each single nanosheet, the reduced symmetry of the monoclinic $VO_2$ lattice when cooling down from the high-temperature rutile phase is expected to induce the formation of four in-plane structural domains [29]. The inverse pole figures shown in Figure 5(f) show two of these because, due to symmetry (presence of a mirror plane) of the monoclinic phase, only half of the complete stereographic projection is displayed. Furthermore, in domain 2, only 1 orientation is seen as we are looking along the $2_1$ axis.

Having proven the excellent crystalline quality and the exquisite control over the orientation of $VO_2$ grown on nanosheets of different types, spectroscopy in the soft X-ray regime is now used to benchmark the epitaxial samples grown on TO nanosheets and provide comparison of the sample quality to what is known in the literature. As shown in the data of Figure 2(c)-(e), the nanosheet approach enables deposition of high-quality $VO_2$ on $Si_3N_4$ membranes that are soft X-ray transparent, opening a route to conducting X-ray spectroscopy in transmission. This yields a bulk-sensitive and direct measure of the absorption that can directly be correlated with local measurements carried out in the TEM. The majority of previous X-ray absorption studies of $VO_2$ have used indirect methods such as Total Electron Yield (TEY) to monitor the X-Ray absorption process [5,30–35].

Linking to the transport data presented in Figure 4, X-ray absorption experiments at the vanadium-$L_{2,3}$ (2p→3d transitions) and oxygen-K (1s→2p transitions) absorption edges also directly probe the MIT, and are readily accessible using soft X-rays provided by a synchrotron light source. In correlated transition metal oxides such as $VO_2$, the ability of soft X-ray spectroscopy to provide detailed information on the manifold and coupled degrees of freedom (e.g. lattice, spin, charge and orbital) using the transition metal-$L_{2,3}$ and O-K edges have been studied extensively [5,30–35]. Panel (a) of **Figure 6** shows V-L$_{2,3}$ edge data both for the metallic (rutile) and insulating (monoclinic) phases for the TO nanosheet templated VO$_2$. As reported by Aetukuri et al.[31], the changes seen to occur across the MIT are related to the orbital occupation, and in particular they underscore the transformation of the three-dimensional rutile situation to one in which V-dimers form along the direction of the rutile c-axis, leading to shifts and splitting of electronic states related to the orbitals polarized in this direction, referred to as the d$_\parallel$ states. For grazing incidence (E⊥c$_{rutile}$) the temperature dependent changes shown in Figure 6(b) agree excellently with published data on films grown on single crystalline, bulk substrates[31], attesting to the quality of the nanosheet templated VO$_2$ thin films. Closer examination of the insets yields that the bulk (transmission) XAS shows an onset of the insulator-metal transition on warming at 340 K, and that ca. 40 K of further heating are required to complete the conversion to the rutile phase. The transport data shown in Figure 4 from fully analogous VO$_2$ films, show an earlier onset and faster completion of the transformation on heating. This can be understood straightforwardly as resulting from the percolative nature of the transport probe on the one hand and to the bulk-sensitive, volume-fraction-driven absorption of soft X-ray radiation on the other hand.

One of the clear order parameters for the MIT is the opening of an energy gap in the insulating phase. This can be clearly seen in XAS at the O-K edge, as shown in Figure 6(b) and panels (b) and (c) of **Figure 7**, and reported in the literature recently[5]. With reference to the electronic structure schematic shown under the data of Figure 6(b), the most marked spectroscopic changes in the O-K edge spectra while entering the metallic phase are due to the closure of the gap, and the disappearance of the unoccupied d$_\parallel$* states, the latter present in the monoclinic phase due to a splitting of the highly directional d$_\parallel$ states. The insets to Figure 6(b) show how these changes to the gap [d$_\parallel$* states] leads to an increase [decrease] of the XAS absorption at the characteristic energy of 529.1 [530.6] eV.

Figure 7(a) illustrates the experimental configuration used for polarized XAS in transmission. For the TO-nanosheet-templated $VO_2$ films, the rutile c-axis is in the film plane, and its in-plane orientation varies from one nanosheet to the next. The synchrotron X-ray beam is large enough to average over a large number of nanosheets, meaning that we can consider vertically polarized radiation (LV) at grazing incidence to yield an unpolarized spectrum. Horizontally polarized radiation (LH) aligns the E-vector perpendicular to the film plane and hence $E \perp c_{rutile}$ is always realized, regardless of the direction of the in-plane orientation of the rutile c-axis in each individual nanosheet-templated epitaxial grain. In Figure 7(b) and 7(c), polarization-dependent XAS spectra are shown for both grazing and normal incidence of the beam, respectively. As is clear from the earlier discussion, for normal incidence (Figure 7(c)), whether linear vertical or linear horizontal radiation is used makes no difference to the absorption spectra, as in all cases a mixture of $E \| c_{rutile}$ and $E \perp c_{rutile}$ is the result. For grazing incidence (Figure 7(b)) and linear horizontal radiation, the E vector is $E \perp c_{rutile}$, compared to mixed $E \perp c_{rutile}$ and $E \| c_{rutile}$ for linear vertical. This is of high relevance for future experiments such as those outlined in **Figure 8** in the next section. The major advantages are the presence of: (i) the O-K leading edge shift at ~529 eV (ii) the directional $d_\|^*$ states in the monoclinic phase at ~530.5 eV and (iii) the $\sigma^*$ states in the rutile phase at ~531.5 eV which all show up clearly in the transmission XAS measurements, also without the need of a full suite of polarization-dependent measurements.

## 3. Conclusion and outlook

This paper reports the successful deposition of high-quality vanadium dioxide thin films on Si substrates and $Si_3N_4$ membranes using oxide nanosheets of $Ti_{0.87}O_2$ (TO) and $NbWO_6$ (NWO) as a templating layer. Each nanosheet is a single crystal template, able to orient the growth direction of the $VO_2$ film as a result of lattice match of the oxygen frameworks of the two systems and compatibility of the 2D atomic structure between the nanosheets and the $VO_2$ crystal planes. X-ray diffraction, SEM imaging, EBSD and ACOM-TEM data all agree on the film orientation, attesting to the epitaxial relationship such that TO nanosheets template $(110)_R$ growth and NWO templates $(002)_R$ growth of $VO_2$. It was also shown that even micron-scale, local control over the $VO_2$ orientation is achievable, when using lithographically patterned nanosheet templates on a single, monolithic substrate.

The second main strand in the research presented deals with the metal-insulator transition of the nanosheet templated $VO_2$ films. Due to strain effects along the *c*-axis of $VO_2$ rutile phase, $T_{MIT}$ was 10 K higher [9 K lower] on $Ti_{0.87}O_2$ [$NbWO_6$] nanosheets, compared to bulk $VO_2$ single crystal values. Truly bulk-sensitive soft X-ray experiments carried out in transmission also underlined the excellent quality of the $VO_2$ thin films, which displayed all the hallmarks of the MIT known from films grown on bulk substrates. The significant soft X-ray transmission contrast – for example at the O-K leading edge, or at the position of $d_\parallel*$ feature – combined with the ability to work in transmission, means such nanosheet templated $VO_2$ films are ideal for advanced soft X-ray techniques such as lens-less imaging of the MIT with spatial resolution of tens of nm. Such techniques involve holographic reconstruction of the real-space patterns formed during the MIT, and can be carried out on a modified version of the samples used for the studies reported here. Figure 8 shows a schematic for such an experiment. We have taken the first step towards such experiments and have recently successfully reconstructed first images during the MIT of $VO_2$ using the HERALDO reconstruction technique [36–38].

To conclude, we have shown that nanosheets can act as templates for heteroepitaxial growth of high-quality $VO_2$ thin films on arbitrary substrates. The latter can be amorphous or have a very different crystallographic structure compared to the target material $VO_2$ itself. This approach allows this important test-case material for oxide-based devices and switching to be grown tailored for specific device applications and for fundamental research. An example is given of how micro-structured areas of differing $VO_2$ orientation can be generated, and how the nanosheet-templated growth approach can be used to enable soft X-ray holographic lens-less imaging of the metal-insulator transition in $VO_2$.

**4. Experimental Section**

*Preparation of nanosheet films*

Potassium carbonate K$_2$CO$_3$ (Fluka), lithium carbonate Li$_2$CO$_3$ (Riedel-de Haen), titanium (IV) dioxide TiO$_2$ (Sigma-Aldrich), niobium (V) oxide Nb$_2$O$_5$ (Alfa Aesar), and tungsten (VI) oxide WO$_3$ (Alfa Aesar) had a purity of 99.0% or higher and were used as received. Nitric acid HNO$_3$ (65%, ACROS Organics) and *tetra*-n-butylammonium hydroxide TBAOH (40% wt. H$_2$O, Alfa Aesar) were used as received. Demineralized water was used throughout the experiments. K$_{0.8}$Ti$_{1.73}$Li$_{0.27}$O$_4$ and LiNbWO$_6$ were synthesized as reported in the literature [39,40], and the layered protonated titanate, H$_{1.07}$Ti$_{1.73}$O$_4$.H$_2$O, and protonated HNbWO$_6$.xH$_2$O were obtained by treating with 2 M HNO$_3$ for 3 days and replacing new acid solution every day. The rapid exfoliation with TBAOH, which has been reported for H$_{1.07}$Ti$_{1.73}$O$_4$.H$_2$O [41] was also observed for HNbWO$_6$.xH$_2$O. Full-coverage TO and NWO nanosheet films were generated on Si substrates and Si$_3$N$_4$ TEM grips using the LB method.

The micron-scale patterned nanosheet template combining TO and NWO nanosheets was prepared as follows. A monolayer of TO nanosheets was deposited on a Si substrate. On top of this monolayer, hexamethyldisiloxane (Merck) was spin-coated at 3000 rpm for 30 s and then a thick layer of photoresist (OiR 907-12 from Olin Microelectronic Materials Inc.) was spin-coated at 3000 rpm for 30 s. After heating at 90°C for 2 min, the sample was exposed to a Hg lamp with a wavelength of 365 nm for 10 s under a 20 µm spaced line grating mask (in a Karl Suss MA56 Mask Aligner). The photoresist was developed for 1 min (in OPD 4262 from Arch Chemicals) and baked at 110°C for 2 min. Subsequently, a monolayer of NWO nanosheets was deposited on this line-patterned sample. Lastly, the sample was dipped and held upside-down in acetone for 1 min for lift-off, in order to remove the NWO nanosheets on top of the photoresist. The sample was then rinsed with ethanol and dried in a N$_2$ gas stream.

*Pulsed laser deposition of VO$_2$ thin films*

Pulsed laser deposition was carried out in a vacuum system equipped with a KrF excimer laser, with a wavelength of 248 nm (COMPEX from Coherent Inc.). The central part of the laser beam was selected with a mask and focused on a polycrystalline $V_2O_5$ target. The deposition conditions of the $VO_2$ films were: laser repetition rate 4 Hz, energy density 1.3 J cm$^{-2}$, spot size 1.8 mm$^2$, oxygen pressure 7.5 mTorr, deposition temperature 520°C, number of pulses 15000, and substrate-target distance 50 mm. After deposition, the samples were cooled down to room temperature at a maximum rate of 5°C min$^{-1}$ at the deposition pressure.

*Analysis and characterization*

The surfaces of the nanosheet films and of the $VO_2$ on Si substrates were investigated using atomic force microscopy (AFM, Bruker Dimension ICON) operating in tapping mode and the data were processed using Gwyddion software (version 2.48) [42]. The relative coverage of nanosheets on substrates was determined at four different locations. The temperature dependent crystal structure of $VO_2$ thin films was analyzed using X-ray diffraction θ-2θ scans (XRD, PANalytical X'Pert Pro MRD) equipped with an Anton Paar DHS 1100 Domed Hot Stage. The temperature dependence of the resistance was measured using a Quantum Design Physical Properties Measurement System. A two-probe measurement was performed to extract the MIT behavior, hysteresis, and transition temperature. High resolution scanning electron microscopy (HR-SEM) and electron backscattering diffraction (EBSD) were performed on a Merlin field emission microscope (Zeiss 1550) equipped with an angle-selective backscatter detector at room temperature. The automated crystal orientation maps in microprobe TEM mode (ACOM-TEM) were acquired on a FEI Tecnai G2 microscope (FEG, 200 kV), equipped with the ASTAR system from Nanomegas. Electron precession was applied to acquire quasi-kinematical data and to facilitate automated indexation. The precession angle used of 0.4° yielded an electron probe size of ~1.5 nm. Interface mapping was achieved in post treatment with the orientation imaging microscopy (OIM) analysis software from EDAX, including noise reduction and texture analysis. Mis-indexed or non-indexed points were corrected using a standard EBSD cleanup procedure. Furthermore, grains smaller than 5 pixels or with a low reliability (<10%) were removed from the analysis. The pole figures were calculated using the harmonic series expansion and were generated along the common directions (001, 110, etc.) with the Gaussian half-width set at 5 degrees. The average orientation was taken from every identified grain.

Soft X-ray transmission experiments were carried out at the UE56-PGM1 beamline at the BESSY II synchrotron source located at the Helmholtz Centre HZB in Berlin. Linearly polarized X-rays (horizontal and vertical) were generated using the UE56 helical undulator, and the energy resolution of the beamline was set to 80 meV. The sample temperature was carefully controlled using a Janis cryostat, and the transmission of X-rays was monitored by comparing photon flux monitors before (refocusing mirror current) and after the sample (on a diode), and included correction for mirror/diode contamination by division with an 'empty scan' in which no $VO_2$ sample is held in the beam path. The soft X-ray lensless imaging exepriments were conducted using the COMET end-station at the SEXTANTS beamline of the SOLEIL synchrotron [43].

**Acknowledgements**

The authors thank Mark A. Smithers for performing high resolution scanning electron microscopy and electron backscattering diffraction. This work is part of the DESCO research programme of the Foundation for Fundamental Research on Matter (FOM), which is part of the Netherlands Organisation for Scientific Research (NWO). P.L. acknowledges the NWO/CW ECHO grant ECHO.15.CM2.043. N.G. acknowledges funding from the *Geconcentreerde Onderzoekacties* (GOA) project "Solarpaint" of the University of Antwerp and the FLAG-ERA JTC 2017 project GRAPH-EYE. G.L. acknowledges financial support from the Flemish Research Fund (FWO) under project G.0365.15N.

Received: ((will be filled in by the editorial staff))
Revised: ((will be filled in by the editorial staff))
Published online: ((will be filled in by the editorial staff))

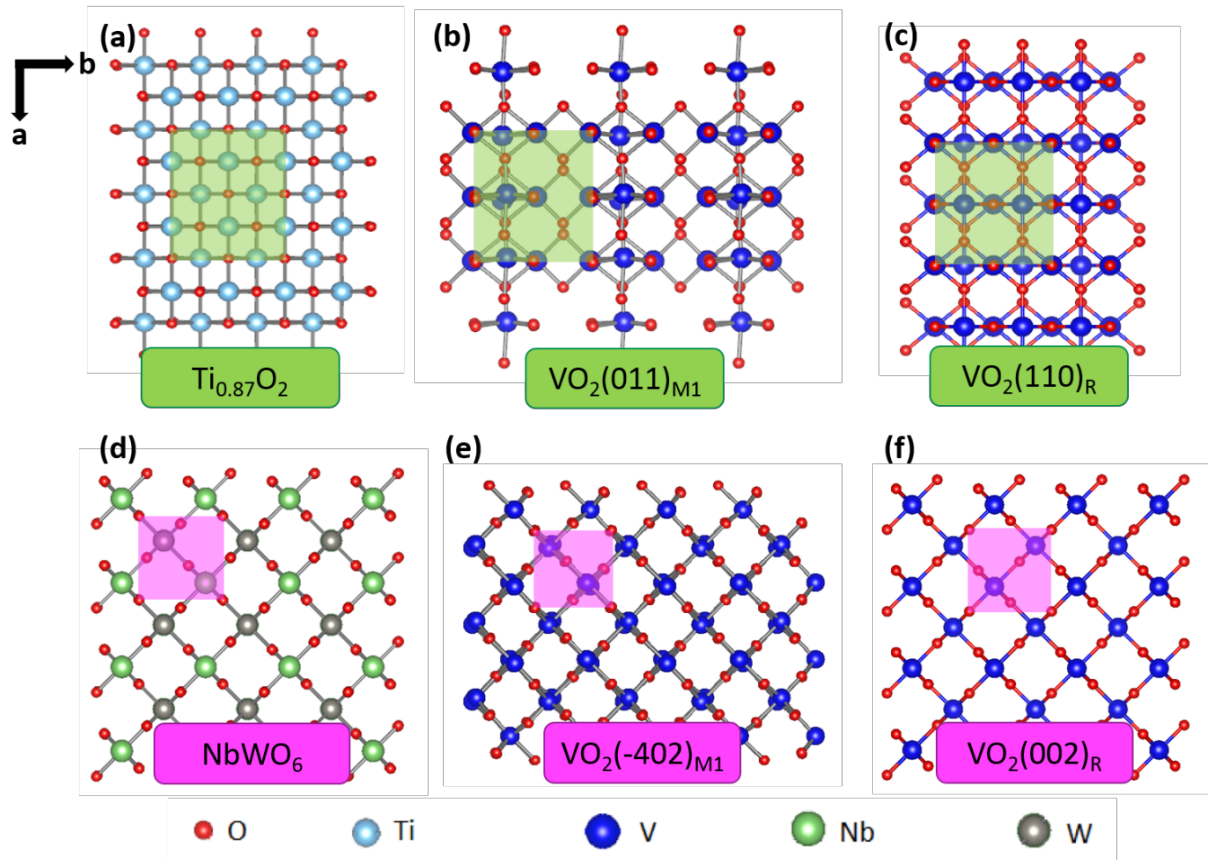

**Figure 1. Excellent epitaxy conditions for nanosheets and VO$_2$.** Schematic illustration of atomic structures for (a) Ti$_{0.87}$O$_2$ nanosheet, (b) VO$_2$ (011)$_{M1}$, (c) VO$_2$ (110)$_R$. A suitable TO unit cell is shown shaded in green. (d) NbWO$_6$ nanosheet, (e) VO$_2$ (-402)$_{M1}$ and (f) VO$_2$ (002)$_R$. A suitable NWO unit cell is shown shaded in pink.

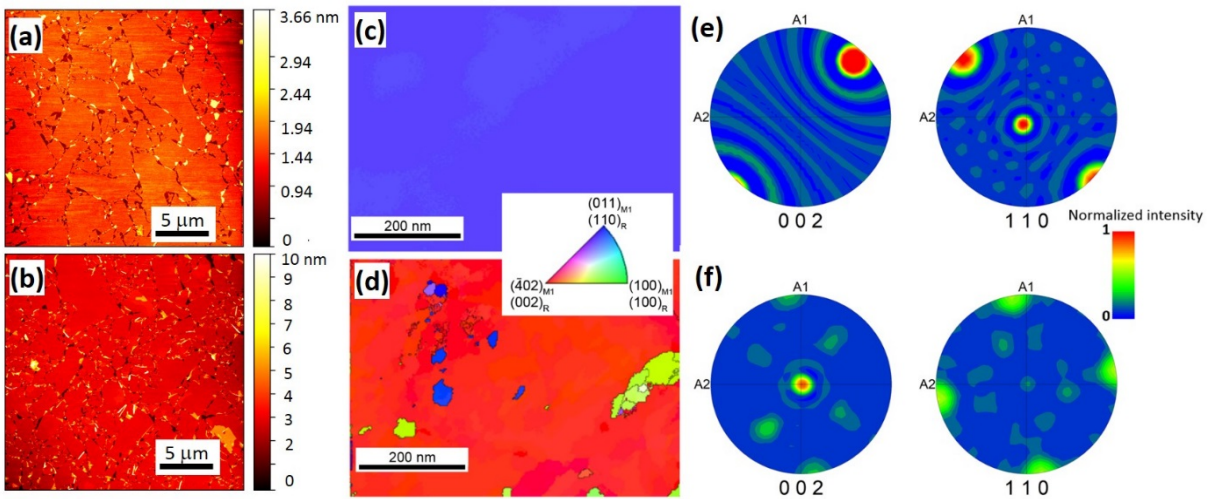

**Figure 2: AFM of nanosheets and automated crystal orientation mapping of VO$_2$ using TEM.** The top [bottom] row of figures is from samples with TO[NWO]-templates. AFM images of the (a) TO and (b) NWO nanosheets showing the homogeneous coverage [scale bar: 5 μm]. (c) and (d) show high temperature ACOM-TEM characterization of the rutile VO$_2$ phase at 423K [scale bar: 200 nm]. (e) Corresponding pole figures for the (002)$_R$ and (110)$_R$ directions, with a color-scaled intensity as shown in the index on the right side. Including is the triangular color scale wedge for the ACOM-TEM data, which also shows the orientational relationship between the $M_1$ and R phases.

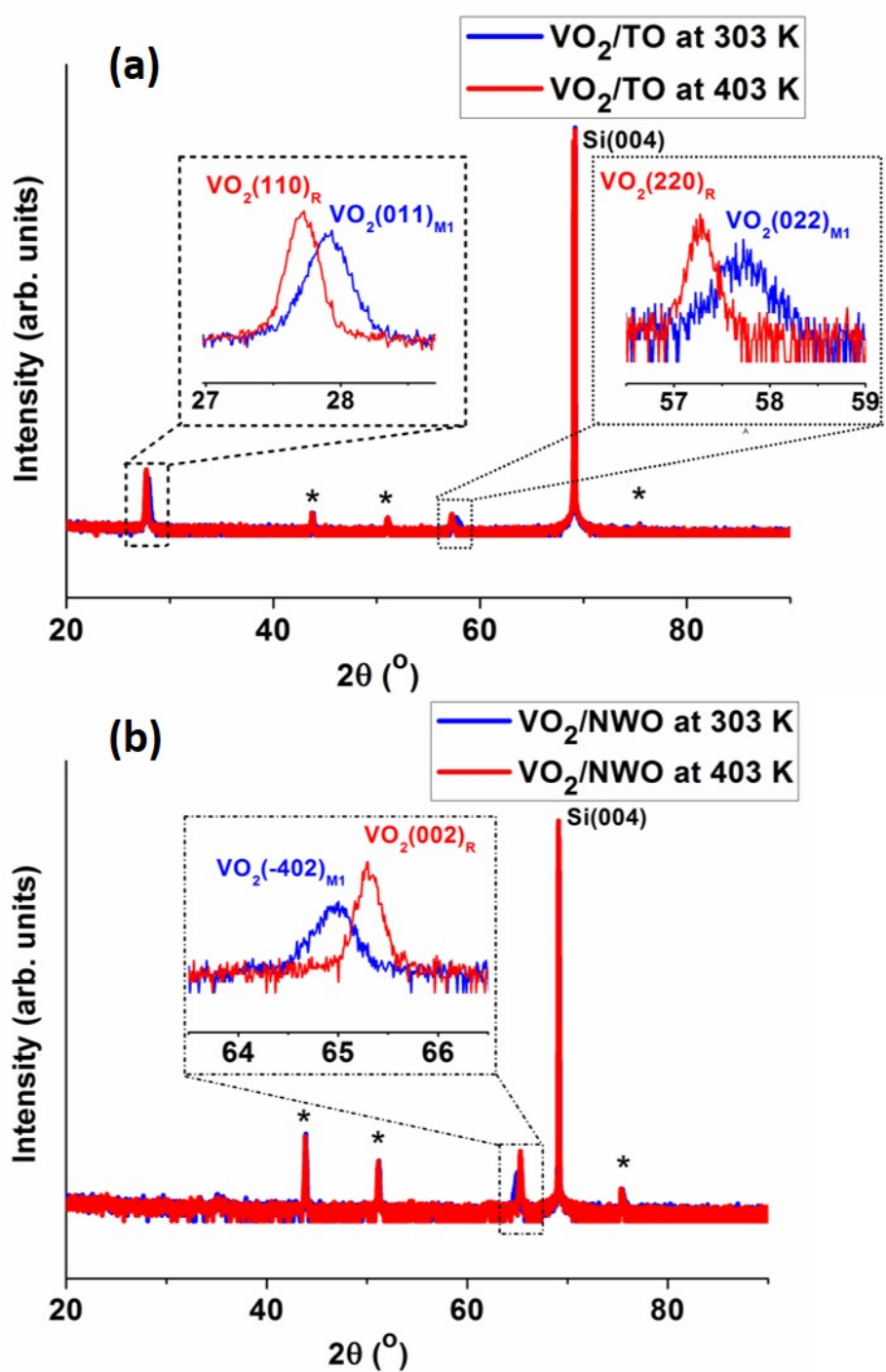

**Figure 3. Temperature-dependent identification of the VO₂ structural phases using X-ray diffraction.** XRD patterns of VO₂ films on (a) TO and (b) NWO nanosheets, measured at 303 K (M1 phase) and 403 K (R phase). The three peaks labelled '*' originate from the Inconel alloy 625/718 clamps holding the sample on the diffractometer heating stage. The VO₂ Bragg peaks can be indexed using the film orientation discussed in the text, fully in agreement with the TEM data of Figure 2 and literature values of the temperature-dependence of the $VO_2$ crystal structure.

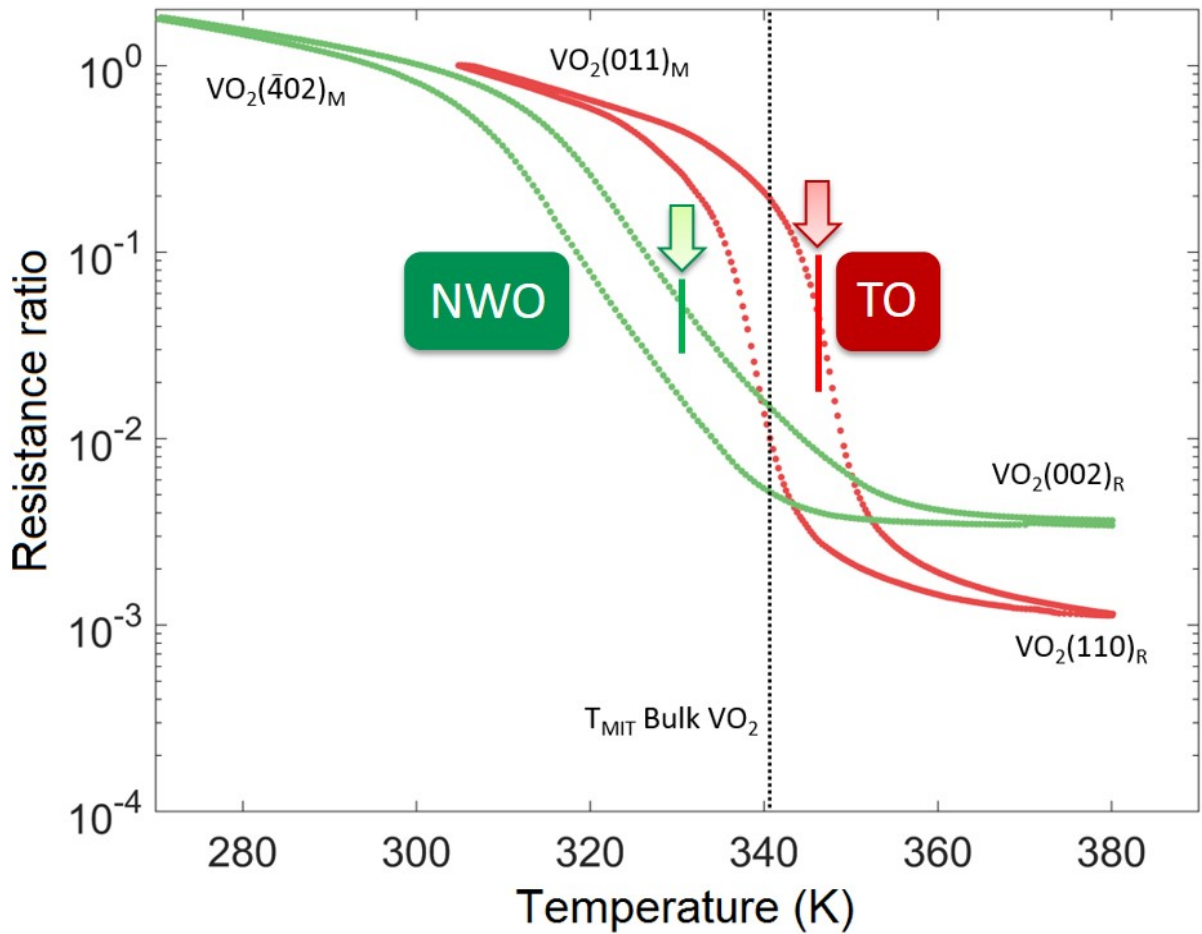

**Figure 4. Transport characterization of the Mott metal-insulator transition of $VO_2$.** Resistance ratio (R[R phase]/R[M1 phase]) of $VO_2$ films grown on TO and NWO nanosheets as a function of temperature. For the M1 phase, the resistance was $2.6 \times 10^5$ Ω [$5.6 \times 10^4$ Ω] for TO [NWO] nanosheet-templated $VO_2$, respectively. The resistance in the $VO_2$-R phase was a factor 810 [280] times lower than in the $M_1$-phase for TO [NWO] nanosheet-templated $VO_2$ growth, respectively.

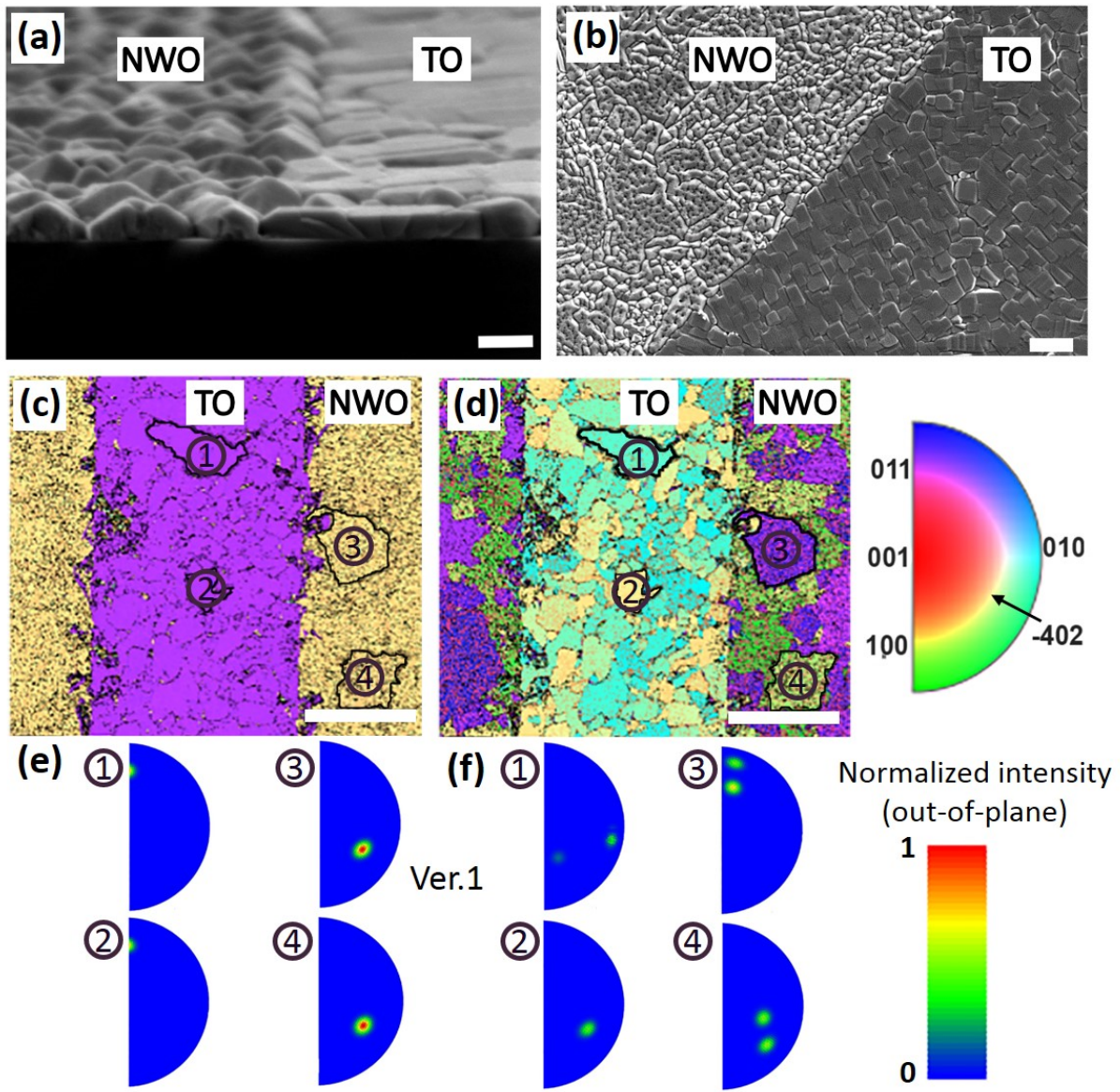

**Figure 5. Micron-level, deterministic control over the VO$_2$ orientation on a single substrate.** Room temperature HRSEM images of the cross-section (a) and plan (b) views at the line boundary of VO$_2$ film (left side is VO$_2$ grown on NWO nanosheet; right side is growth on TO nanosheet). (c) and (d) show inverse pole figure maps, measured using EBSD at room temperature, displaying the VO$_2$ film orientation in the out-of-plane (c) and in-plane (d) directions. In panel (c), the left- and right-side of the image shows growth on NWO nanosheet (gold color). Here, the film normal is the monoclinic (-402) axis (→(002)$_R$). The central, purple-colored strip represents growth on TO nanosheet where the film normal is (011)$_M$, equivalent to (110)$_R$. Panel (d) shows the in-plane texture of the VO$_2$ films along the horizontal direction (x). The scale bars represent 100 nm in (a), 500 nm in (b) and 10 μm in both (c) and (d). Panel (e) shows out-of-plane inverse pole figures corresponding to the four different regions marked on panel (c). The out-of-plane orientation is clearly controlled by the type of nanosheets used. Panel (f) displays in-plane inverse pole figures corresponding to the same four regions marked in panel (d). Four different domains on each individual nanosheet result from the reduced monoclinic symmetry compared to the cubic rutile case. Only half of the stereographic projection is shown due to symmetry resulting in only two spots.

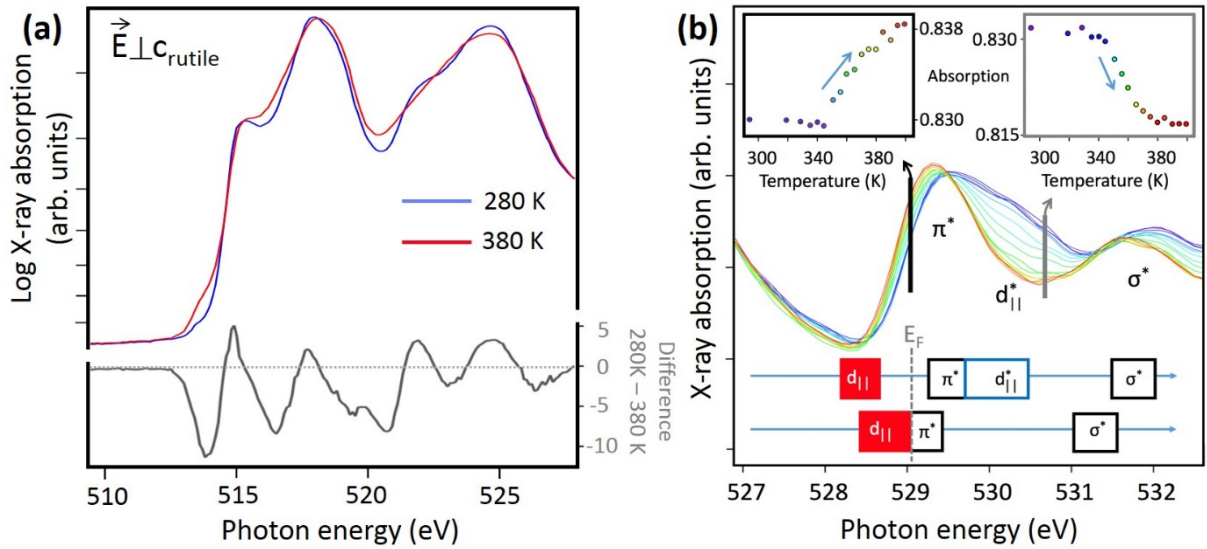

**Figure 6. Soft X-ray absorption in transmission.** V-$L_{2,3}$ XAS of nanosheet-supported $VO_2$ recorded in transmission at the temperatures shown for grazing incidence of the linearly polarized X-rays. Subtle yet clear differences in the spectral features mirror alterations in the electronic structure, reflecting changes in orbital energies as the V-$d_∥$ states split and an energy gap opens in the insulating, low temperature phase. The details of the difference spectrum agree very well with published polarization-dependent V-$L_{2,3}$ X-ray absorption data from $VO_2$ grown epitaxially on bulk single crystalline substrates [31]. Panel (b) shows the O-K edge recorded at normal incidence as a function of temperature. The insets - whose data-points are color-coded to match the spectra from which they are taken - show the T-dependence of the two main absorption features signaling the insulator to metal transition. Increasing leading edge intensity (black arrow/inset) tracks the closing of the insulating gap as the rutile phase is reached, and a different aspect of the same physics yields to the decrease of the $d_∥^*$ feature at 530.6 eV (grey arrow/inset). The identity of the different peaks, together with a schematic representation of the corresponding density of states is given under the data of panel (b).

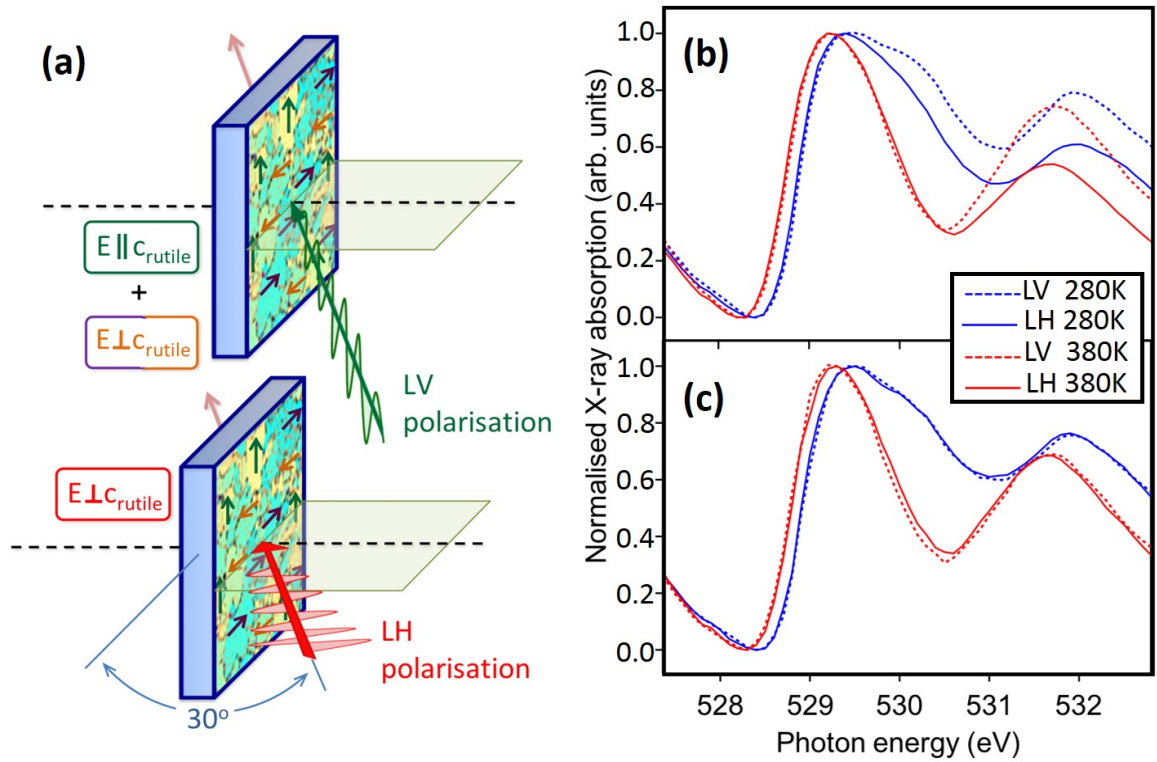

**Figure 7. V orbital occupancy across the MIT. (a)** Schematic of soft X-ray absorption experiments on TO-nanosheet-supported $VO_2$ thin films grown on 20 nm thick silicon nitride TEM windows. Linearly polarized synchrotron radiation impinges in grazing incidence, as indicated (the transmitted beam is measured using a diode downstream of the sample, not shown). The c-axis of the $(110)_R$-$VO_2$ film is oriented differently in each of the nanosheet domains, but is always in the plane of the film. Therefore, linear horizontal (LH) fixes $E \perp c_{rutile}$ and vertical polarization probes a mix of $E \perp c_{rutile}$ and $E \| c_{rutile}$. Polarization dependent measurements at the O K-edge above and below the transition for both grazing [panel (b)] and normal [panel (c)] incidence show the in-plane polarization of the unoccupied portion of the highly directionally aligned $d_\|$ states.

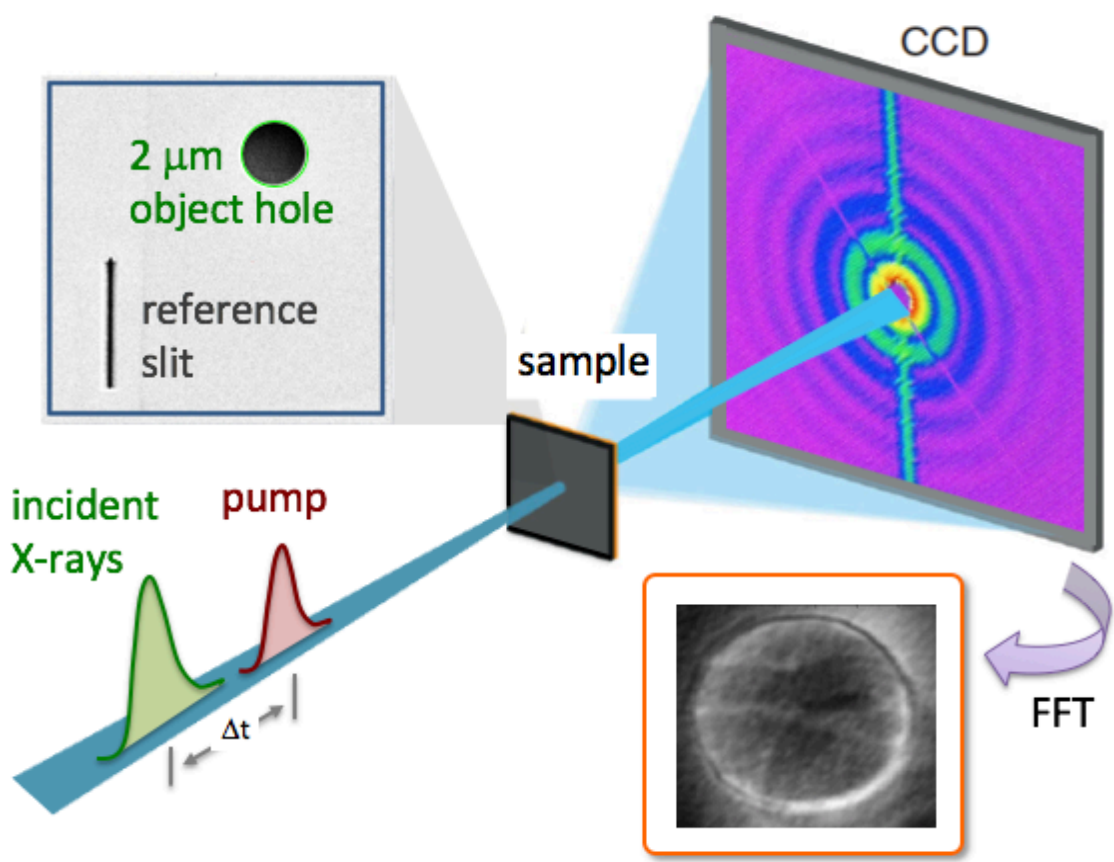

**Figure 8. Lens-less imaging of the MIT of VO₂.** Schematic of a soft X-ray holography experiment on nanosheet-supported VO₂ thin films which incorporate with a gold mask structure. The VO₂ films can be grown using PLD on TO nanosheets that are deposited on commercial, 200 nm thick silicon nitride TEM windows (e.g. Simpore). A gold mask is subsequently deposited and a sample window (diameter 2μm) + reference slit are machined into the mask using a focused ion beam (FIB). The window reveals the VO₂ film, whereas the slit goes right through the whole structure (including the VO₂, the nanosheets and silicon nitride, too). Illumination with coherent X-rays yields a far-field diffraction pattern, and using a differential filter and fast Fourier transform it is possible to reconstruct a real space image of the different phases of VO₂ during the MIT. At an X-ray free electron laser, sufficient intensity in a single ultrafast flash of X-rays would also allow a pump-probe version of this experiment, so enabling a stop-motion film to be built up of how VO₂ switches on both the fs timescale and nm length scales.

**Table 1. Symmetry and lattice constants of TO and NWO nanosheets.**

| Nanosheet | 2D structure | Lattice constant (Å) |
|---|---|---|
| $Ti_{0.87}O_2^{0.52-}$ | Rectangular | $a$ = 3.76 ; $b$ = 2.97 |
| $NbWO_6^-$ | Square | $a$ = 4.68 |

**Oxide heteroepitaxy of VO$_2$ on technical substrates** is realized using pulsed laser deposition. VO$_2$ films are grown in (110)$_R$ and (002)$_R$ orientation using Ti$_{0.87}$O$_2$ and NbWO$_6$ nanosheet templates on Si and Si$_3$N$_4$ membranes. Local control of VO$_2$ orientation on the micrometer level on a single, arbitrary substrate is demonstrated, and soft X-ray experiments on the metal-insulator transition are conducted in transmission.

62 words

**Keyword** nanosheets, vanadium dioxide, X-ray absorption, transmission, lens-less imaging

Corresponding authors:    G.Koster@utwente.nl

Nicolas.Gauquelin@uantwerpen.be

M.S.Golden@uva.nl

**Title**
**Tailoring Vanadium Dioxide Film Orientation using Nanosheets: A Combined Microscopy, Diffraction, Transport and Soft X-ray in Transmission Study**

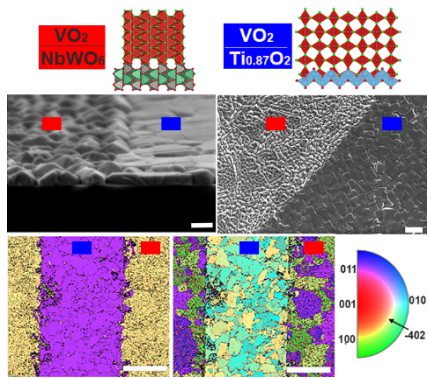